\documentclass[aps,twocolumn,prx,superscriptaddress,letterpaper]{revtex4-1}
\usepackage{amssymb}
\usepackage{mathtools}
\usepackage[pdftex]{graphicx}
\usepackage[usenames, dvipsnames, svgnames, table]{xcolor}
\usepackage[colorlinks=true, citecolor=Blue,
            linkcolor=BrickRed, urlcolor=ForestGreen]{hyperref}
\usepackage{txfonts}
\usepackage{mathrsfs}
\usepackage{bm}
\usepackage{multirow}
\begin{document}

%%%%%% Some definition %%%%%%

\newcommand{\be}{\begin{equation}}
\newcommand{\ee}{\end{equation}}
\newcommand{\R}[1]{\textcolor{red}{#1}}
\newcommand{\B}[1]{\textcolor{blue}{#1}}
\newcommand{\fixme}[1]{\textcolor{orange}{#1}}

%%%%%% Title %%%%%%

\title{Quantum Correlation of Light Mediated by Gravity}

\author{Haixing Miao} 
\email{haixing@star.sr.bham.ac.uk}
\affiliation{School of Physics and Astronomy, and Institute for 
Gravitational Wave Astronomy, 
University of Birmingham, Edgbaston, Birmingham B15 2TT, United Kingdom}
\author{Denis Martynov}
\email{dmartynov@star.sr.bham.ac.uk}
\affiliation{School of Physics and Astronomy, and Institute for Gravitational Wave Astronomy,   
University of Birmingham, Edgbaston, Birmingham B15 2TT, United Kingdom}
\author{ Huan Yang }
\email{hyang@perimeterinstitute.ca}
\affiliation{Perimeter Institute for Theoretical Physics, Waterloo, ON N2L2Y5, Canada}
\affiliation{University of Guelph, Guelph, ON N2L3G1, Canada}
\author{Animesh Datta}
\email{animesh.datta@warwick.ac.uk}
\affiliation{Department of Physics, University of Warwick, Coventry CV4 7AL, United Kingdom}

%%%%%% Abstract %%%%%%

\begin{abstract}
We propose to explore the quantum nature of gravity using
the correlation of light between 
two optomechanical cavities, and the quantumness of the 
correlation is witnessed by squeezing. 
As long as the gravity between the
end mirrors of two cavities is quantum in the Newtonian limit, 
we show that the squeezing is always nonzero and monotonically increases
as the mechanical property of the mirrors is improved.  The proposed scheme 
provides a new pathway for 
 testing the quantum nature of gravity systematically with
tabletop experiments. \end{abstract}

\maketitle

%%%%%% Introduction %%%%%%

{\it Introduction --- }Constructing a consistent and verifiable quantum 
theory of gravity has been a longstanding challenge of modern 
physics\,\cite{Kiefer2006, Woodard2009, Oriti2009}, which 
is partially due to the difficulty in experimentally observing quantum 
effects of gravity. This, to certain extents, motivates some theoretical models that treat gravity as a fundamental 
classical entity\,\cite{Moller1962, Roenfeld1963, Kibble1978, Adler2007, Carlip2008, Yang2013a, Anastopoulos2014, Bahrami2014} or 
being emerged from yet unknown underlying microphysics\,\cite{Jacobson1995, Verlinde2011, Padmanabhan2015, Hossenfelder2017}. 
Experimentally probing the quantum nature of gravity is therefore essential for 
providing hints towards constructing the correct model\,\cite{Howl2018, Carney2019}. 
Recently, two experimental proposals have been made to demonstrate gravity-induced quantum entanglement between two mesoscopic test masses\,\cite{Bose2017, Marletto2017}, motivated by an early
suggestion of Feynman\,\cite{Feynman1957}. Both involve two matter-wave interferometers located close to each other such that their test masses can be entangled through the gravitational interaction.  
Whether gravity-mediated entanglement in the Newtonian limit establishes the quantumness of gravity or not has been debated\,\cite{Krisnanda2017, Anastopoulos2018, Hall2018, Belenchia2018b, Reginatto2018}, because the radiative degrees of freedom---the graviton, are not directly probed in these experiments. Nonetheless, such experiments are important steps towards understanding gravity in the quantum regime\,\cite{Diosi1987, Penrose1996, Bassi2013, Helou2017, Vinante2017, Bassi2017}. 

\begin{figure}[b]
\includegraphics[width=\columnwidth]{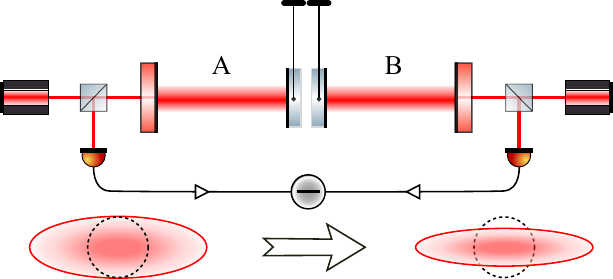}
\caption{Schematics showing the proposed experimental setup 
(the local oscillator for the
 homodyne detection of the outgoing field of B is not shown). 
Squeezing of the outgoing field of A conditional on 
measuring that of B manifests the gravity-mediated quantum
 correlation. Dashed circle denotes the vacuum level. }
\label{fig:config}
\end{figure}

The challenge of demonstrating gravity-induced entanglement is achieving a very low thermal decoherence rate, and is beyond what can be achieved with the state-of-the-art instruments, as illustrated in the Appendix\,\ref{app:ent_cond}.
In this paper, we propose a tabletop optomechanical experiment to explore gravity-mediated quantum correlation of light. The strength of the correlation 
is quantified by squeezing, which is non-classical according to 
the Glauber-Sudarshan distribution function\,\cite{Glauber1963, Sudarshan1963, Titulaer1965}.
The setup is shown schematically 
in Fig.\,\ref{fig:config}. Two optomechanical cavities
are placed close to 
each other with their end mirrors interacting 
through gravity. In contrast to the single-photon nonlinear regime studied by Balushi {\it et al.}\,\cite{Balushi2018}, we consider the linear regime with the cavity driven by a coherent laser field. The quantum correlation is inferred by squeezing of the outgoing field of 
the cavity A conditional on the homodyne measurement of the outgoing field of B. 

If the gravitational interaction between two mirrors
is quantum in the Newtonian limit, namely, 
\begin{equation}\label{eq:newton}
\hat H_{AB} = - \frac{G m_A m_B}{|\hat q_A-\hat q_B|}\,, 
\end{equation}
we will show such a conditional squeezing is always nonzero. 
Observing a sizeable squeezing however
requires the optomechanical cavities to be
quantum radiation pressure limited, in which case the 
squeezing can be approximately as
\begin{align}\label{eq:sqz}\nonumber
{\cal S} &= 10\log_{10} \left[1 + \left( \frac{2 Q_m G\rho}{\omega_m^2}\right)^2 \right] \\ &\approx  2\,{\rm dB} \left(\frac{0.5\,\rm Hz}{\omega_m/2\pi}\right)^4 \left(\frac{Q_m}{3\times 10^6}\right)^2 \left(\frac{\rho}{20\,\rm g/cm^3 }\right)^2\,.
\end{align}
It only depends on the gravitational constant $G$, material density 
$\rho$, mechanical frequency $\omega_m$, and quality factor $Q_m$. 

The statistical uncertainty of the measurement will affect the squeezing signal. 
Fortunately, because the system is in a steady state, the signal-to-noise ratio (SNR) increases as
the measurement time $\tau$. Achieving a unity SNR requires 
\begin{equation}\label{eq:tau_est}
\tau \approx 1\,{\rm year}\left(\frac{\omega_m/2\pi}{0.5\,\rm Hz}\right)^3 \left(\frac{3\times10^6}{Q_m}\right) \left(\frac{20\,\rm g/cm^3 }{\rho}\right)^2\,. 
\end{equation} 
Both $\cal S$ and $\tau$ scale rapidly with $\omega_m$, and low-frequency 
mechanical oscillators are therefore preferable. 

There are several optomechanical experiments that have achieved
the quantum radiation pressure limited regime
but with high-frequency mechanical oscillators\,\cite{Purdy2013, Moeller2017, Rossi2018, Cripe2018, Barzanjeh2018, Delic2020} and in particular, Ref.\,\cite{Barzanjeh2018} reported 
a steady-state entanglement between light mediated by 
a mechanical oscillator. Advancing these
experimental techniques towards low frequencies, also an effort in the gravitational-wave community\,\cite{Punturo10a, Rana:Review2013, abbott2017exploring, Yu2017}, is the key to measure the 
gravity-mediated quantum correlation. 

{\it Dynamics --- }The derivation of Eq.\,\eqref{eq:sqz} 
follows the linear-dynamics analysis
in quantum optomechanics\,\cite{Chen2013, Aspelmeyer2014}: Solving 
the linear Heisenberg equations of motion for dynamical variables, which 
are the mirror position and quadratures of 
the outgoing optical fields, and 
representing them in terms of external fields, which are 
the ingoing optical fields 
and the thermal bath field. 

The total Hamiltonian of the system is $\hat H_{\rm tot} =\hat H_A+\hat H_B
+\hat H_{AB}$. The individual cavity is quantified by the 
standard linearised optomechanical Hamiltonian, 
which describes the radiation-pressure coupling between the 
optical field and the 
centre of mass motion of the 
mirror (mechanical degree of freedom). The 
interaction part of $\hat H_A$ for cavity A is (similarly for B):
\begin{equation}\label{eq:H_optint}
\hat H_{A}^{\rm int}  =  \hbar\,\omega_q \hat X_A \hat Q_A\,.
\end{equation}
We denote $\hat X_A$ as the amplitude quadrature of the cavity mode, which 
is conjugate to the phase quadrature $\hat Y_A$: $[\hat X_A,\,\hat Y_A]=i$, and 
$\hat Q_A$ as the
mirror position $\hat q_A$ normalised with respect to its
zero-point motion $\sqrt{\hbar/(2m \omega_m)}$. 
The parameter $\omega_q$ 
describes the optomechanical coupling strength:
\begin{equation}
\omega_q \equiv \sqrt{\frac{2P_{\rm cav}\omega_0}{m c L \omega_m}}\,, 
\end{equation}
which depends on the  intra-cavity optical power $P_{\rm cav}$, 
the laser frequency $\omega_0$, the mirror  
mass $m$, and the cavity length $L$. 
 
Up to the second-order of the mirror position, the non-trivial interaction part of  
$\hat H_{AB}$ in Eq.\,\eqref{eq:newton} is 
\begin{equation}\label{eq:Hgrav}
\hat H_{AB}=  \hbar
\frac{\omega^2_g}{\omega_m} \hat Q_A \hat Q_B\,.
\end{equation}
Here we have assumed two mirrors 
having the same mechanical frequency and mass $m_A=m_B=m$.
The characteristic gravitational interaction frequency 
$\omega_g$ is equal to $\sqrt{G m/d^3}$ when the two mirrors have 
a mean separation $d$ much larger than their size, which is the case for 
mesoscopic levitating masses considered in Refs.\,\cite{Bose2017, Marletto2017, Qvarfort2018, Delic2020}. For macroscopic test mass mirrors of gram or kilogram scale,
their separation can be made comparable to their size (yet not affected by e.g.
the Casimir force), and we have 
\begin{equation}\label{eq:omegag}
\omega_g = \sqrt{\Lambda G \rho}\,, 
\end{equation}
which does not explicitly depend on the mirror mass. 
The form factor  
$\Lambda$ is determined by the geometry of two mirrors. It 
is $\pi/3$ for two spheres with the mean separation equal to 
twice of the radius, and we assume $\Lambda = 2.0$ throughout the paper, which is a good approximation
for two closely-located disks with 
the radius being 1.5 times its thickness (see Appendix \ref{app:geometry} for details).

Solving the Heisenberg equations of motion results in the 
following frequency domain input-output relation for cavity A (similarly for cavity B):
\begin{align}
\hat X^{\rm out}_A(\omega) & = \hat X_A^{\rm in}(\omega)\,,\\
\hat Y^{\rm out}_A(\omega) & = \hat Y_A^{\rm in}(\omega)+\sqrt{2/\gamma}\,\omega_q
\hat Q_A(\omega)\,, 
\end{align}
where we have assumed that the cavity bandwidth $\gamma$ is much 
larger than the frequency of interest so that the 
cavity mode can be adiabatically eliminated, cf. Eq.\,(2.68) of Ref.\,\cite{Chen2013}. 
The position of mirror  
A satisfies 
\begin{equation}\label{eq:QA}
\hat Q_A = \chi_{qq}
[\sqrt{\gamma/2}\,\omega_q\hat X^{\rm in}_A-
(\omega_g^2/\omega_m) \hat Q_B +2\sqrt{\gamma_m}\hat Q^{\rm th}_A]\,.
\end{equation}
Here
$\chi_{qq}\equiv -\omega_m/(\omega^2-\omega_m^2+i\gamma_m \omega)$ 
is the susceptibility with the mechanical damping rate $\gamma_m\equiv \omega_m/Q_m$; $\hat Q^{\rm th}$ is the normalised
thermal Langevin force according to the fluctuation-dissipation theorem\,\cite{Callen1951, Kubo1966}, and its double-sided 
spectral density is equal to $\bar n_{\rm th} + (1/2)$ with the thermal occupation 
number $\bar n_{\rm th}\equiv k_B T/(\hbar \omega_m)$
in the high-temperature limit.

\begin{figure}[b]
\includegraphics[width=\columnwidth]{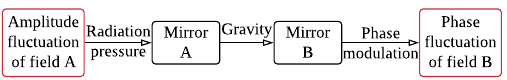}
\caption{A flow chart illustrating the physical meaning of $\cal G$ introduced in 
the input-output relation Eq.\,\eqref{eq:io}.}
\label{fig:gfactor}
\end{figure} 

The final input-output relation involving both cavities is 
\begin{equation}\label{eq:io}
\left[\begin{array}{c} 
\hat X^{\rm out}_A \\ \hat  Y_A^{\rm out} \\ 
\hat  X^{\rm out}_B \\ \hat  Y_B^{\rm out}\end{array}\right] = \left[ \begin{array}{cccc} 
1 & 0 & 0 & 0 \\
{\cal K} & 1 & {\cal G} & 0 \\
0 & 0 & 1 & 0 \\
{\cal G} & 0 & {\cal K} & 1 
\end{array}\right] 
\left[\begin{array}{c} 
\hat  X^{\rm in}_A \\ \hat Y_A^{\rm in} \\ 
\hat  X^{\rm in}_B \\ \hat Y_B^{\rm in}\end{array}\right] +
 \left[ \begin{array}{cc} 
0 & 0 \\
\alpha & \beta  \\
0 & 0  \\
\beta & \alpha
\end{array}\right] 
\left[\begin{array}{c} 
\hat Q_A^{\rm th}  \\ 
\hat Q_B^{\rm th} \end{array}\right]\,.
\end{equation}
Here ${\cal K}\equiv -4\omega_q^2 \chi_{qq}/\gamma$ quantifies the correlation between the amplitude quadrature and the phase quadrature in the individual cavity and is responsible for the
optomechanical squeezing\,\cite{Kimble02, Brooks2012, Safavi-Naeini2013a, Purdy2013b, Buchmann2016}. The two parameters 
$\alpha\equiv 2\sqrt{2\gamma_m/\gamma}\,\omega_q \chi_{qq}$ and $\beta\equiv \alpha \chi_{qq} (\omega_g^2/\omega_m)$ quantify the output response  to the thermal 
force noise. As illustrated in Fig.\,\ref{fig:gfactor}, the dimensionless parameter $\cal G$ quantifies the mutual correlation between two cavities and is defined as ${\cal G}\equiv {4 \omega_q^2 \omega_g^2 \chi_{qq}^2}/({\gamma \omega_m})$. Its magnitude reaches the maximum at the mechanical frequency:
\begin{equation}
|{\cal G}(\omega_m)|   = 
 2\,{\cal C}\, Q_m \left(\frac{\omega_g}{\omega_m}\right)^2\,. 
\end{equation}
The optomechanical 
cooperativity defined as 
\begin{equation}
{\cal C} \equiv \frac{2\omega_q^2}{\gamma \gamma_m}
\end{equation}
is proportional to the number of intra-cavity photons\,\cite{Aspelmeyer2014}. The fact that $|\cal G|$ is proportional to ${\cal C}$ shows that the optomechanical interaction coherently enhances the correlation by amplifying the quantum fluctuation of light.  

{\it Quantum correlation and conditional squeezing --- }Notice that the correlation reaches the maximum around the mechanical  frequency within a narrow frequency bandwidth defined by $\gamma_m$. We can therefore focus on the
quadratures of the outgoing fields around $\omega_m$ with a bandwidth $\Delta \omega$ comparable to $\gamma_m$ (or the measurement time comparable to the damping time $\tau_m \equiv 2\pi Q_m/\omega_m$). The corresponding normalised quadrature operators 
are defined as 
\begin{equation}
\hat {\cal X} \equiv \sqrt{\Delta \omega/\pi}\, \hat X^{\rm out}(\omega_m)\,, 
\quad 
\hat {\cal Y} \equiv \sqrt{\Delta \omega/\pi}\, \hat Y^{\rm out}(\omega_m)\,. 
\end{equation}
They satisfy $[\hat {\cal X},\, \hat {\cal Y}^{\dag}]=2i$, where we have 
approximated the Dirac delta function $\delta(0)$ as $1/\Delta\omega$.  
With such a normalisation, the uncertainty of $\hat {\cal X}$ 
or $\hat{\cal Y}$ for the vacuum or coherent
state is equal to 1. 

Due to the quantum correlation, the uncertainty of the amplitude quadrature of
A can be reduced after we measure the phase
quadrature of B.  
 The conditional uncertainty 
is obtained by minimising the residue over the filtering function $\cal F$: 
\begin{align}\nonumber
\sigma^{\rm cond}_{\cal XX} & = \min_{\cal F} {\rm Tr}\left[\hat \varrho \left(\hat{\cal X}_A - {\cal F}\hat  {\cal Y}_B \right)^2 \right]= 
\sigma_{\cal XX}- \frac{|\sigma_{\cal XY}|^2}{\sigma_{\cal YY}} 
\\\label{eq:condv}
& = 1 -\frac{|{\cal G}|^2}{1+|{\cal K}|^2 + |{\cal G}|^2 + (2\bar n_{\rm th}+1)(|\alpha|^2+|\beta|^2)}\,, 
\end{align}
where we define the variance $\sigma_{\cal XX}\equiv {\rm Tr}[\hat \varrho \, \hat {\cal X}_A \hat {\cal X}_A^{\dag} ]$ (similar for $\sigma_{\cal YY}$ 
of $\hat{\cal Y}_B$), and the covariance $\sigma_{\cal XY}\equiv  {\rm Tr}[\hat \varrho\, (\hat {\cal X}_A \hat {\cal Y}_B^{\dag} + \hat {\cal Y}_B^{\dag}\hat {\cal X}_A )/2]$ with $\hat \varrho$ being the density matrix. 
In obtaining the above result, we 
have used the fact that the ingoing optical field is in the vacuum state because the coherent amplitude is absorbed by the coupling rate $\omega_q$\,\cite{Chen2013, Aspelmeyer2014}.
The corresponding optimal Wiener filter is given by ${\cal F}_{\rm opt} = \sigma_{\cal XY}/\sigma_{\cal YY}= {\cal G}/[1+|{\cal K}|^2 + |{\cal G}|^2 + (2\bar n_{\rm th}+1)(|\alpha|^2+|\beta|^2)]$. 

\begin{figure}[t]
\includegraphics[width=\columnwidth]{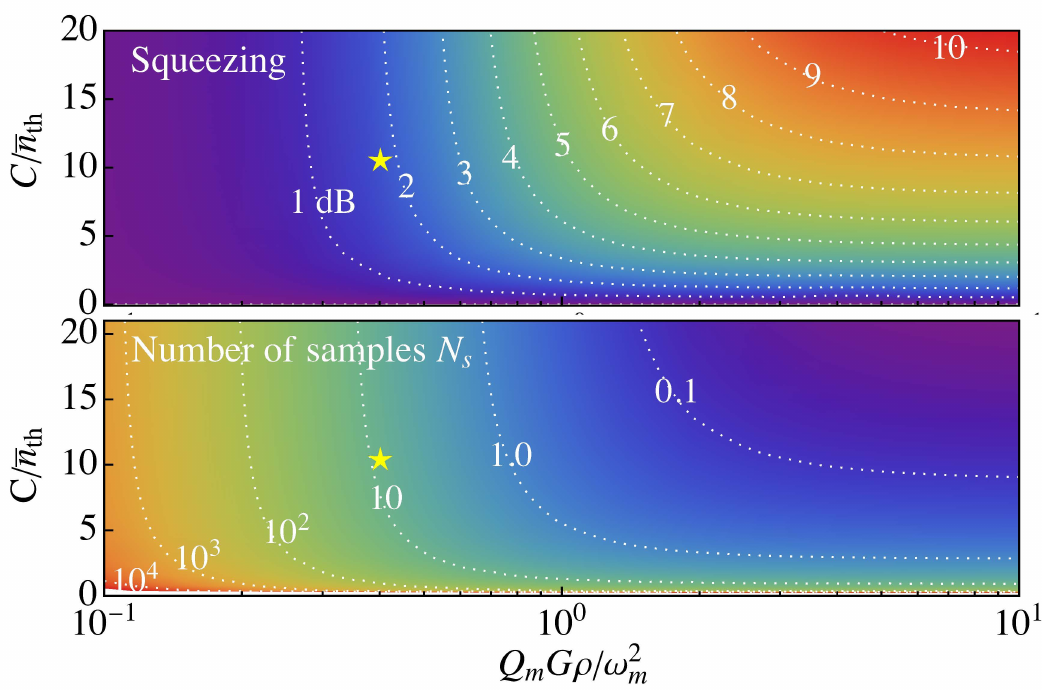}
\caption{The top panel shows the squeezing in dB as a function of two dimensionless parameters: ${\cal C}/{\bar n_{\rm th}}$ and $Q_m G\rho/\omega_m^2$. The bottom panel shows the minimum $N_s$
needed to achieve a unity signal-to-noise ratio ($N_s<1$ implies 
one sample is sufficient), and a small $N_s$ does not mean a short 
measurement time, which is equal to $N_s$ times the mechanical 
damping time. The two stars on the graphs mark the parameters
 assumed in Eq.\,\eqref{eq:sqz} and Eq.\,\eqref{eq:tau_est} of the 
 introduction part.}
\label{fig:sqz}
\end{figure}

As we can see from Eq.\,\eqref{eq:condv}, 
the conditional uncertainty of $\hat {\cal X}_A$ is always smaller than 1, 
which implies squeezing. To observe such a conditional squeezing experimentally, the estimation error due to a finite number of measurements needs to be smaller than the squeezing level. 
According to the standard estimation theory, the 
unbiased estimator for the conditional uncertainty for a known average is
\begin{equation}
\sigma_{\cal XX}^{\rm est} = \frac{1}{N_s}
\sum_{k=1}^{N_s}\tilde \sigma^{\rm cond}_{\cal XX}(k)\,,
\end{equation}
where $\tilde \sigma^{\rm cond}_{\cal XX}(k)$ is the conditional variance
for the $k$-th measurement sample and $N_s$ is the total number of samples.
In our case, each sample corresponds to a measurement time of 
the order of the mechanical damping time $\tau_m$. 
For a total measurement time of $\tau$, we have 
\begin{equation}\label{eq:Ns}
N_s \equiv  \frac{\tau}{\tau_m} = \frac{\omega_m \tau}{2\pi Q_m}\,. 
\end{equation}
Since $\sum_{k=1}^{N_s}\tilde \sigma^{\rm cond}_{\cal XX}(k)$ follows the 
chi-squared distribution with $N_s$ degrees of freedom, the estimation error 
is equal to $\sqrt{{2}/{N_s}}\, \sigma^{\rm cond}_{\cal XX}$. 
It needs to be smaller than the squeezing level to achieve
a unity SNR, which implies
\begin{equation}\label{eq:snr_cond}
\sqrt{ \frac{2}{N_s}}\,\sigma^{\rm cond}_{\cal XX} \leq \frac{|{\cal G}|^2}{1+|{\cal K}|^2 + |{\cal G}|^2 + (2\bar n_{\rm th}+1)(|\alpha|^2+|\beta|^2)} \,. 
\end{equation} 
The above condition leads to a requirement on the minimum measurement time $\tau$. 
For experimentally relevant parameters, we have $\bar n_{\rm th}\gg 1$ and $|{\cal K}|\gg 1$, we can approximate the denominator of Eq.\,\eqref{eq:condv} and 
Eq.\,\eqref{eq:snr_cond} 
as $|{\cal K}|^2 + |{\cal G}|^2 + 2\bar n_{\rm th}(|\alpha|^2+|\beta|^2)$. 
The resulting squeezing and also the minimum number of samples
are shown in Fig.\,\ref{fig:sqz}. They only depend on two characteristic dimensionless parameters: ${\cal C}/{\bar n}_{\rm th}$, the ratio between the optomechanical cooperativity and the thermal occupation number, and $Q_m G\rho/\omega_m^2$, solely determined by the gravity and the mechanical property of the mirror. 

To obtain a sizeable squeezing, we learn from Fig.\,\ref{fig:sqz} that first 
$Q_m G\rho/\omega_m^2$ needs to be large, which implies high-quality-factor,
low-frequency test mass mirrors, and second the cooperativity shall be
 much
larger than the mean thermal occupation number, namely, 
\begin{equation}
{\cal C} \gg \bar n_{\rm th}\,. 
\end{equation}
This corresponds to the quantum radiation pressure limited 
regime in optomechanics\,\cite{Aspelmeyer2014}. In such a regime, 
the squeezing and minimum number of samples turn out to become independent of the optical property and only depend on the mechanical property. In particular, we have 
\begin{equation}
\sigma_{\cal XX}^{\rm cond} \approx \frac{1}{1+|{\cal G}/{\cal K}|^2} = 
\frac{1}{1+ (2 Q_m { G\rho}/{\omega_m^2})^2}\,, 
\end{equation}
which, written in terms of dB, 
gives rise to Eq.\,\eqref{eq:sqz} shown in the introduction. 
The minimum number of samples $N_s$ 
to achieve a unity SNR can be approximated as 
\begin{equation}
N_s \approx 1 + 4|{\cal K}/{\cal G}|^2
\approx \left(\frac{\omega_m^2}{ Q_m G \rho }\right)^2\,. 
\end{equation}
The second approximation is satisfied for those parameter values assumed in Eq.\,\eqref{eq:tau_est} where we have shown the equivalent minimum measurement time. 

{\it Conclusions and discussions --- }To summarise, our approach for probing the 
quantum nature of gravity takes advantage of new advancements in quantum 
optomechanical experiments. It  
is complimentary to other approaches based upon matter-wave interferometers. 
In general, achieving a sizeable squeezing requires quantum radiation 
pressure limited systems with high-quality-factor, low-frequency mechanical 
test mass mirrors. Even though 
the squeezing signal does not explicitly depend on the size of the test mass mirror, 
having a low mechanical frequency usually implies macroscopic test masses. 
For illustration, we provide a possible 
set of sample parameters to reach ${\cal C}/\bar n_{\rm th}$ of the order of 10 
 implicitly assumed in Eq.\,\eqref{eq:sqz} 
for $\omega_m/(2\pi) = 0.5\,\rm Hz$ and $Q_m=3\times 10^6$:
\begin{equation}
\frac{\cal C} {\bar n_{\rm th}}\approx 10 \left(\frac{1\,\rm g}{m}\right) \left(\frac{P_{\rm cav}}{2\,{\rm kW}}\right)\left(\frac{\rm Finesse}{4000}\right)\left(\frac{300\,\rm K}{T}\right)\,,
\end{equation}
which corresponds to a suspended high-finesse cavity with a gram-scale 
test mass mirror at room temperature, close to what has been achieved by the MIT group\,\cite{Corbitt2007}. The gravity experiments with milligram test masses\,\cite{Schmole2016, Matsumoto2018}
can be promising if pushed to the low-frequency regime.

Let us consider the consequence of 
different outcomes of the measurement that we propose. If we do not detect a
predicted level of squeezing after a careful calibration of the system, it will imply that the assumption on the gravity sector is invalid, cf. Eq.\,\eqref{eq:newton}, as the quantum aspects of the optomechanical interactions have already been established experimentally. One compelling possibility then is that gravity is classical, 
so that it does not appear in the quantum interaction Hamiltonian. If we do observe a
non-zero squeezing, we will be able to rule out classical models of gravity, 
in particular the Schr\"odinger-Newton (SN)
type of classical gravity models---the gravity is sourced by the 
expectation value of quantum matters\,\cite{Moller1962, Roenfeld1963, Kibble1978, Adler2007, Carlip2008, Yang2013a, Anastopoulos2014, Bahrami2014}, 
which does not lead to quantum correlation.  
This is because the corresponding SN
 two-body interaction for the optomechanical setup would be, cf. Eq.\,(27) of Ref.\,\cite{Yang2013a}, 
\begin{equation}
\hat H_{\rm AB}^{\rm SN} = \hbar \frac{\omega_g^2}{2\omega_m} \left(\langle \hat Q_A\rangle  \hat Q_B + \hat Q_A \langle \hat Q_B\rangle\right) \,.
\end{equation}
According to Eq.\,\eqref{eq:QA}, the quantum part of 
$\langle \hat Q_A\rangle$ or $\langle \hat Q_B\rangle $ is zero, 
as the expectation value of the quantum fluctuation $\hat X_A^{\rm in}$ 
is zero. For future study, it would be interesting also to explore the predictions of
emergent gravity models\,\cite{Jacobson1995, Verlinde2011, Padmanabhan2015, Hossenfelder2017} on the conditional squeezing level in this
proposed optomechanical setup. 

{\it Acknowledgements --- }We would like to thank Yanbei Chen, 
Chris Collins, Bassam Helou, and Dominic Branford
for fruitful discussions, and Joe Bentley for proofreading the manuscript.
 H.M. is supported by UK STFC Ernest Rutherford 
Fellowship (Grant No. ST/M005844/11). 
H.Y. is supported by the Natural Sciences and Engineering Research 
Council of Canada, and by Perimeter Institute for Theoretical Physics. 
D.M. acknowledges the support from the Institute for Gravitational-wave 
Astronomy at University of Birmingham. A.D. is supported, in part, by the
UK EPSRC (EP/K04057X/2), and the UK National
Quantum Technologies Programme (EP/M01326X/1,
EP/M013243/1). 

\appendix

\section{Condition for realising gravity-mediated entanglement}
\label{app:ent_cond}

Here we try to derive the general condition for achieving entanglement between the 
outgoing fields of two cavities. The entanglement measure 
can be derived from their total covariance matrix 
 ${\bm \sigma}\equiv {\rm Tr}\{\hat \varrho\,[\hat {\cal X}_A\; \hat {\cal Y}_A\; \hat {\cal X}_B\;\hat {\cal Y}_B]^{\rm T}  [\hat {\cal X}^{\dag}_A\; \hat {\cal Y}^{\dag}_A\; \hat{\cal X}^{\dag}_B\;\hat {\cal Y}^{\dag}_B]\}_{\rm sym}$ where the superscript ``T" means transpose and the subscript ``sym" means symmetrisation: ${\rm Tr}[\hat \varrho\,\hat {\cal X} \hat {\cal Y}]_{\rm sym}\equiv{\rm Tr}[\hat \varrho\,(\hat {\cal X} \hat {\cal Y}^{\dag} + \hat {\cal Y}^{\dag}\hat {\cal X} )/2]$, more explicitly, 
\begin{equation}\label{eq:covmat}
{\bm \sigma}\equiv \left[\begin{array}{cccc}
{\bm \sigma}_A & {\bm \sigma}_{AB} \\
{\bm \sigma}^{\rm T}_{AB} & {\bm \sigma}_B
\end{array}\right]\,.
\end{equation}
The diagonal components ${\bm \sigma}_A = {\bm \sigma}_B $ are
\begin{equation}\label{eq:VA}
{\bm \sigma}_A = \left[\begin{array}{cc} 1 & {\cal K}^* \\ {\cal K} & 
1+|{\cal K}|^2 + |{\cal G}|^2 + (2\bar n_{\rm th}+1)(|\alpha|^2+|\beta|^2)\end{array}\right]\,. 
\end{equation}
The off-diagonal one, describing the cross correlation, is 
\begin{equation}\label{eq:VAB}
{\bm \sigma}_{AB} = \left[\begin{array}{cc} 0 & {\cal G} \\ {\cal G} & 0\end{array}\right]\,. 
\end{equation} 
All the above quantities ${\cal K}$, $\cal G$, $\alpha$ and $\beta$ are referring to their values at  $\omega_m$, in particular, 
\begin{equation}
{\cal K}(\omega_m)= -2 i {\cal C}\,,\quad \alpha(\omega_m)= 2 i \sqrt{\cal C}.
\end{equation} 
Note that ${\cal K}(\omega_m)$ is complex and it leads to the complex squeezing, 
which is unaccessible with the standard homodyne detection\,\cite{Purdy2013b, Buchmann2016}. That is why the noise ellipse of A illustrated in Fig.\,\ref{fig:config} shows no correlation between the amplitude quadrature and 
the phase quadrature of A. 

The figure of merit for quantifying such a bipartite Gaussian 
entanglement is the so-called logarithmic negativity ${\cal E}_{N}$\,\cite{Simon2000, Horodecki2009}, which is defined as
\begin{equation}
{\cal E}_N = {\rm max}\left\{-(1/2)\ln\left[\left(\Sigma-\sqrt{\Sigma^2-4{\det \bm\sigma}}\right)/2\right] ,\, 0\right\}\,, 
\end{equation}
where $\Sigma \equiv \det {\bm \sigma}_A +\det {\bm \sigma}_B -2\,\det {\bm \sigma}_{AB}$. A nonzero 
${\cal E}_{N}$ implies the existence of entanglement. In our case, the first term is equal to
\begin{equation}\label{eq:logN}
-\ln\left[\sqrt{1+|{\cal G}|^2+(2\bar n_{\rm th}+1)(|\alpha|^2+|\beta|^2)}-|{\cal G}|\right]\,.
\end{equation}
Having it larger than zero requires 
\begin{equation}\label{eq:ent_cond1}
(2\bar n_{\rm th}+1)(|\alpha|^2 +|\beta|^2) < 2|{\cal G}|\,.
\end{equation}
When using the fact that $|\alpha|\gg |\beta|$ and $\bar n_{\rm th}\gg 1$, we arrive at the  following condition:
\begin{equation}\label{eq:ent_cond}
\gamma_m k_B T \leq \hbar G \rho\,.
\end{equation}
As an order of magnitude, it implies
\begin{equation}\label{eq:ent_cond0}
\frac{T}{Q_m} \leq 3.0\times 10^{-18} {\rm K}\,\left(\frac{0.5\,\rm Hz}{\omega_m/2\pi}\right)\left(\frac{\rho}{20\,\rm g/cm^3}\right)\,.
\end{equation}
This requirement is  beyond what we can achieve with the state-of-the-art
instruments, and needs further experimental efforts.
Note that a related analysis 
of steady-state Gaussian entanglement in the case of 
two levitating nanobeads has also been presented by Qvarfort {\it et al.}\,\cite{Qvarfort2018}.

The above requirement Eq.\,\eqref{eq:ent_cond} turns out to be equally applicable to the free-mass case with the resonant frequency $\omega_m\rightarrow 0$, as $\omega_m$ does not appear explicitly in the equation. We consider the standard thermal decoherence model. The corresponding master equation for the density 
matrix $\hat \varrho$ of 
the two test masses takes the following diffusive form:
\begin{equation}
\dot {\hat \varrho}(t) = \frac{i}{\hbar}\left[\hat \varrho(t),\,\hat H_{AB}\right] - \frac{2 m \gamma_m k_B T\delta x_q^2}
{\hbar^2} \sum_{j=A, B} \left [\hat Q_j,\, \left [\hat Q_j,\, \hat \varrho(t)\right ]\right]\,, 
\end{equation}
where $\delta x_q$ is the characteristic length scale and is 
equal to the Standard 
Quantum Limit (SQL)\,\cite{Braginsky92} for Gaussian states and
the size of 
the quantum superposition for non-Gaussian states.
For the quantum entanglement 
to survive in the presence of the thermal decoherence, we require the interaction rate 
to be larger than the decoherence rate:
\begin{equation}\label{eq:ent_req}
\frac{||\hat H_{AB}||}{\hbar}\ge \frac{2 m \gamma_m k_B T\delta x_q^2}
{\hbar^2}\,,
\end{equation} 
where $||\hat H_{AB}||$ is the norm that quantifies the magnitude of the
gravitational-interaction energy when A and B are at the quantum level. 

In the case of $\delta x_q$ much smaller than the mean separation $d$, we 
have, according to Eq.\,\eqref{eq:Hgrav}, 
\begin{equation}\label{eq:HAB_gaussian}
||\hat H_{AB}|| \approx 2 \Lambda G m \rho \delta x_q^2\,,
\end{equation}
where we have assumed that  $\delta x_q$ is the same for A and B. 
The condition Eq.\,\eqref{eq:ent_req} 
leads to Eq.\,\eqref{eq:ent_cond} for $\Lambda$ being the order of 1. 
Similarly, when $\delta x_q$ is much larger than the mean separation $d$, 
e.g. the non-Gaussian superposition state in the setup using 
 the matter-wave interferometers\,\cite{Bose2017, Marletto2017}, the corresponding gravitational 
interaction energy is simply 
\begin{equation}\label{eq:HAB_non_Gaussian}
||\hat H_{AB}|| =\frac{G m^2}{d}\,. 
\end{equation}
 Eq.\,\eqref{eq:ent_req} results in 
\begin{equation}\label{eq:req_non_Gaussian}
\gamma_m k_B T \leq \frac{\hbar G m}{2 d \delta x_q^2} < \frac{\hbar G m}{2 d^3} \leq \hbar G\rho\,,
\end{equation}
where in the last inequality we have used the fact that $m/d^3$ is at most of the order of 
the matter density $\rho$. 
Therefore, regardless whether the two test masses (either being free mass or harmonic oscillator) are prepared in Gaussian states or non-Gaussian states, the same requirement universally applies for achieving the gravity-mediated entanglement in the presence of thermal decoherence. 

\section{Dependence of $\Lambda$ on the test mass geometry}
\label{app:geometry}
\begin{figure}[t]
\includegraphics[width=\columnwidth]{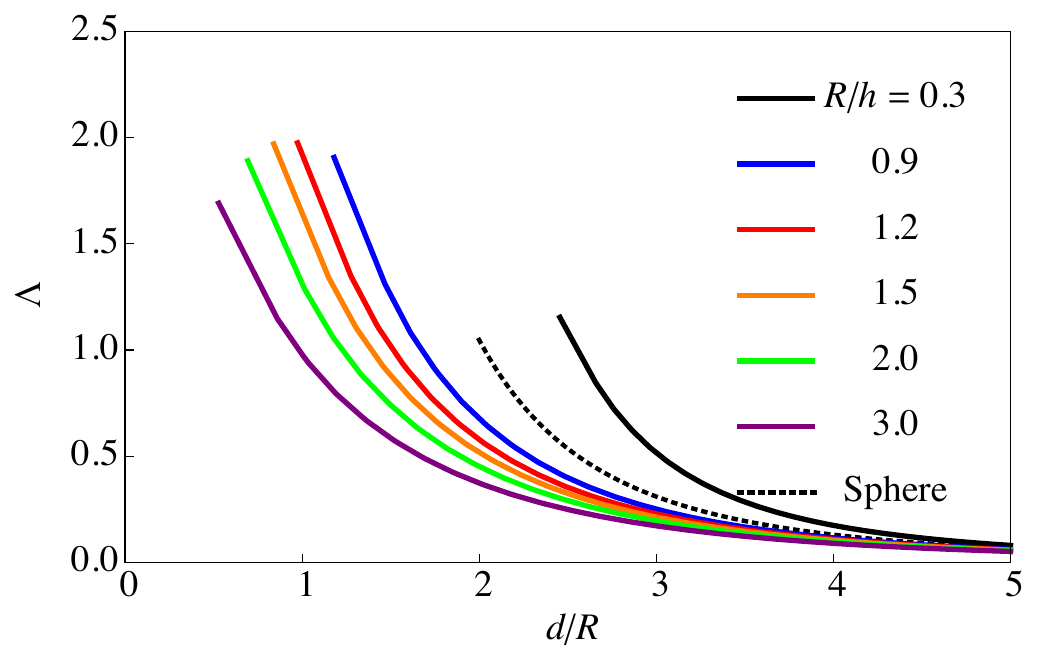}
\caption{The form factor $\Lambda$ as a function of distance for different ratios between 
the radius $R$ and the thickness $h$ of the disk. As a reference, we also show the case of two 
spheres in dashed line. The lower bound of the distance for different curves are defined by 
the one when the two disks touch each other.}
\label{fig:Lambda}
\end{figure}

Depending on the geometry of the two test masses, the form factor in defining $\omega_g$ in Eq.\,\eqref{eq:omegag} is different. The simplest case is having two identical spheres with a uniform density, 
and $\Lambda = \pi/3$ when their mean separation is equal to twice of their radius. Here we consider 
two test masses that have the shape of a disk which is usually the geometry for mirrors of optical cavities. 
Since there is no analytical expression for the Newtonian force between two disks, we perform numerical 
integration of the force for disks with different ratios between the radius $R$ and the thickness 
$h$. We then take the derivative numerically 
with respect to their mean separation $d$ along the optical axis to obtain $\Lambda$ for different mean separations and the maximum $\Lambda$ is achieved when their surfaces are close to each other with $d$ approximately equal to $h$. Fig.\,\ref{fig:Lambda} shows the result, and we can see 
the maximum value of $\Lambda$ for $R/h=1.5$ is around 2.0, which is the one we assumed in the main text. 

%%%%%% References %%%%%%

\bibliography{references}

%%%%%% End %%%%%%

\end{document}